\documentclass[10pt]{article}

\usepackage{graphicx}

\usepackage{natbib}
\usepackage{amsmath}
\DeclareMathOperator{\Cov}{Cov}
\DeclareMathOperator{\Var}{Var}

\usepackage[margin=2cm]{geometry}

\begin{document}

\title{Exploration of an Interdisciplinary Scientific Landscape}

\author{Juste Raimbault$^{1,2}$\medskip\\
$^1$ UMR CNRS 8504 G{\'e}ographie-cit{\'e}s  \\
              $^2$ UMR-T IFSTTAR 9403 LVMT\medskip\\
\texttt{juste.raimbault@polytechnique.edu}
}

\date{}

Raimbault, J. (2019). Exploration of an interdisciplinary scientific landscape. Scientometrics, 119(2), 617-641.

{\let\newpage\relax\maketitle}

\begin{abstract}
Patterns of interdisciplinarity in science can be quantified through complementary dimensions. This paper studies as a case study the scientific environment of a generalist journal in Geography, \emph{Cybergeo}, in order to introduce a novel methodology combining citation network analysis and semantic analysis. We collect a large corpus of around 200,000 articles with their abstracts and the corresponding citation network that provides a first citation classification. Relevant keywords are extracted for each article through text-mining, allowing us to construct a semantic classification. We study the qualitative patterns of relations between endogenous disciplines within each classification, and finally show the complementarity of classifications and of their associated interdisciplinarity measures. The tools we develop accordingly are open and reusable for similar large scale studies of scientific environments. Our contribution therefore provides, besides the methodology, a new way to construct open databases and study journals for which data are difficult to obtain.
\medskip

\end{abstract}

\section*{Introduction}
\label{sec:intro}

The development of interdisciplinary approaches is increasingly necessary for most of disciplines, both for further knowledge discovery but also societal impact of discoveries, as it was for example recently coined by the special issue of Nature~\citep{natureInterdisc}. Interdisciplinary research has furthermore a higher citation impact as shown by \cite{CHEN20151034}, but traditional research institutions such as the Nobel price still fail to foster interdisciplinarity~\citep{szell2018nobel}. \cite{banos2013pour} suggests that the development of such approaches must occur within a subtle spiral between and inside disciplines. An other way to interpret this phenomenon is to read it as the emergence of vertically integrated fields conjointly with horizontal questions as detailed in the Complex Systems roadmap \citep{2009arXiv0907.2221B}. There are naturally multiple views on what exactly interdisciplinarity is and it actually depends on the domains involved: recent hybrid disciplines (see e.g. the ones exhibited by \cite{bais2010praise} such as astro-biology) are a good illustration of the case where entanglement is strong and new discoveries are vertically deep, whereas more loose fields such as ``urbanism'', which have no precise definition and where integration is by essence horizontal, is an example of how transversal knowledge can be produced. Interactions between disciplines are not always smooth, as shows the misunderstandings when urban issues were recently introduced to physicists as \cite{dupuy2015sciences} recalls.

\subsection*{Quantitative studies of knowledge}

These concerns are part of an understanding of processes of knowledge production in which evidence-based perspectives, involving quantitative approaches, play an important role. These paradigms can be understood as a \emph{quantitative epistemology} following \cite{chavalarias2013phylomemetic}. Quantitative measures of interdisciplinarity would therefore be part of a multidimensional approach of the study of science that is in a way ``beyond bibliometrics''~\citep{cronin2014beyond}. The focus of this paper is positioned within this stream of research.

The possible methods for quantitative insights into epistemology are numerous. A good illustration of the variety of approaches is given by network analysis. Using citation network features, a good predicting power for citation patterns is for example obtained by~\cite{2013arXiv1310.8220N}. The visualisation of citation networks can help to identify qualitative changes in the structure of disciplines \citep{chen2004searching}. Visual analytics indeed play a crucial role in the study of knowledge \citep{borner2003visualizing}.

Co-authorship networks can also be used for predictive models ~\citep{2014arXiv1402.7268S}. Disciplines can be stratified into layers to reveal communities between them and therein collaboration patterns~\citep{2015arXiv150601280B}. Networks analysis are used in other fields such as economics of innovation: for example, \cite{choi2014patent} proposes a method to identify technological opportunities by detecting important keywords from the point of view of topological measures. In political science, \cite{gaumont2017methods} uses the gargantext platform to offer a clear view of political debates seen from the proxy of twitter.

Such works allow to understand the dynamics of knowledge. For example, \cite{shibata2008detecting} use topological analysis of the citation network to detect emerging research fronts. Citation and co-citations patterns are at the basis of the generic framework developed by \cite{chen2006citespace} to detect emerging trends in science. The case of Nanoscience is studied by \cite{bonaccorsi2010proliferation} as a case of an emerging field exhibiting proliferating patterns for the keywords used.

Large scale initiatives to map science, such as the UCSD map of science \citep{borner2012design}, are crucial both for the understanding of scientific environments, for scientific policies, and for social and technological transfers. \cite{leydesdorff2009global} describe a global map with nested discipline maps.

An other important quantitative insight into science is the modeling of its dynamics \citep{borner2011modeling,scharnhorst2012models}, aimed at understanding processes through which collective scientific intelligence and its structures do emerge. This can include the modeling of social processes \citep{edmonds2011simulating}, of research institutions~\citep{Rouse12582}, or of the dynamics of competing theories themselves~\citep{Akerlof13228}.

We now review with more details the different approaches to define and measure interdisciplinarity.

\subsection*{Defining and measuring interdisciplinarity}

Definitions of interdisciplinarity itself and indicators to measure it have already been tackled by a large body of literature. \cite{huutoniemi2010analyzing} recall the difference between \emph{multidisciplinary} (an aggregate of works from different disciplines) and \emph{interdisciplinary} (implying a certain level of integration) approaches. They construct a qualitative framework to classify types of interdisciplinarity, and for example distinguish empirical, theoretical and methodological interdisciplinarities.

\cite{wagner2011approaches} find that knowledge integration is crucial for interdisciplinarity, and that it can occur at different levels, from the single scientist to the research team or the field. \cite{hall2008collaboration} confirm the role of the social aspect in potential interdisciplinary collaborations, by proposing to include the readiness to collaborate in the evaluation of corresponding research units. The multidimensionnal aspect of interdisciplinarity is confirmed even within a specific field such as literature~\citep{austin1996defining}.

Beside these different conceptual approaches to interdisciplinarity, there exist several methodological means to measure it, of which we now give an overview. A first way to quantify interdisciplinarity of a set of publications is to look at the proportion of disciplines outside a main discipline in which they are published, as~\cite{rinia2002impact} do for the evaluation of projects in physics, complementary with judgement of experts. \cite{porter2007measuring} designate this measure as \emph{specialization}, and compare it with a measure of \emph{integration}, also called the \emph{Rao-Stirling} index, which is given by the spread of citations done by a paper within the different Subject Categories (classification of the Web of Knowledge).\cite{lariviere2010relationship} use it on a Web of Science corpus to show the existence of an optimal intermediate level of interdisciplinarity for the citation impact within a five year window. A similar work is done in~\citep{lariviere201410}, focusing on the evolution of measures on a long time range. The influence of missing data on this index is studied by \cite{moreno2016uncertainty}, providing an extended framework taking into account uncertainty.

Other indices based on citation practices have been introduced, such as by \cite{rodriguez2017disciplinarity} which proposes an entropy-based index to classify the behavior of journal regarding knowledge import or export. \cite{zhang2016diversity} recall mathematical properties that need to be verified by diversity indices (such as symmetry or scale-invariance), and proposes that \emph{Hill-type} indices are more relevant than entropy.

\cite{mugabushaka2016bibliometric} show that most of existing indices are a particular cases of \emph{Leinster-Cobbold diversity indices}, which correspond to a third generation of indices to measure diversity in ecology. \cite{leydesdorff2011indicators} compare different indices at the level of journals, and show that they appear to capture different dimensions of this phenomenon in a complementary way.

The use of bottom-up network measures has also been proposed to quantify interdisciplinary research: \cite{porter2009science} combine the integration index with a mapping technique which consists in visualisation of synthetic networks constructed by co-citations between disciplines. \cite{leydesdorff2007betweenness} shows that the betweenness centrality is a relevant indicator of interdisciplinarity, when considering appropriate citation neighborhood. A multilayer network approach was proposed in~\cite{omodei2017evaluating}, using bipartites networks of papers and scholars, in order to produce measures of interdisciplinarity using generalized centrality measures. \cite{rafols2009diversity} combine diversity indices with measures of network coherence, which capture the integration of knowledge.

\subsection*{Semantic analysis and interdisciplinarity}

A particular entry to the quantification of interdisciplinarity is semantic analysis of documents. \cite{nichols2014topic} uses Latent Dirichlet Allocation topic modeling to characterize interdisciplinarity of awards in particular sciences. \cite{palchykov2016ground} do the same for papers in physics based on concept extraction from full texts, and show that the endogenous classes differ from the top-down subjects classification. Semantic networks are otherwise well studied in social sciences, such as for example \cite{2015arXiv151003797G} which analyze semantic networks of political debates. \cite{bouveyron2016stochastic} introduce a block model for network clustering which includes textual information of nodes. \cite{10.1371/journal.pone.0018029} benchmark several text-based clustering methods.

Semantic analysis can be coupled with other dimensions, and in particular citation network analysis. \cite{bras2017oncology} study oncology by coupling the semantic aspects with institutional aspects. \cite{zhang2010journal} describe cross-journals citation clusters in terms of their semantic content, but does not produce an endogenous semantic classification. \cite{gerow2018measuring} finds that the citation influence and the discursive influence of scholars are complementary dimensions of academic success, by using topic modeling for the semantic classification of their work.

\subsection*{Proposed approach and case study}

We develop in this paper a case study coupling citation network exploration and analysis with text-mining, aiming at mapping the scientific landscape in the neighborhood of a particular journal. Our aim is twofold: (i) introduce a methodology coupling the two aspects; (ii) introduce tools which allow the analysis of journals for which data is difficult to obtain, when e.g. not referenced in main databases.

We study an electronic journal in Geography, namely \textit{Cybergeo}\footnote{\texttt{https://journals.openedition.org/cybergeo/}}, that publishes articles within all subfields of Geography and is in that way multidisciplinary. The choice is initially due to data availability, but ensures several constraints making it highly relevant to the objectives given above. First of all, the ``discipline'' of Geography is very broad and by essence interdisciplinary~\citep{bracken2016interdisciplinarity}: the spectrum ranges from Human and Critical geography to physical geography and geomorphology, and interactions between these subfields are numerous. Secondly, bibliographical data is difficult to obtain, raising the concern of how the perception of a scientific landscape may be shaped by actors of the dissemination and thus far from objective, and making technical solutions as the ones we will consequently develop here crucial tools for an open and neutral science. Finally it makes a particularly interesting case study as the editorial policy is generalist and concerned with open science issues such as peer-review ethics transparency~\citep{10.1371/journal.pone.0147913}, open data and model practices, as recalled by~\cite{pumain2015adapting}, and this work contributes to these by fostering the opening of reflexivity.

Our approach combine semantic communities analysis with citation network to extract features such as interdisciplinarity measures. Our contribution is original regarding previous works quantifying interdisciplinarity since it uses bottom-up community reconstruction both in the citation and in the semantic dimension. Our work naturally differ from studies in which the classification mapped is exogenous, such as institutional thesaurus as mapped by \cite{boyack2017thesaurus}. \cite{zhang2010journal} combines citation cluster with semantic analysis, but the semantic clusters are exogenous. We also differ from several previous work using endogenous semantic information, as \cite{vugteveen2014dynamics} for example only uses titles and combine them with references for the proximity measure, whereas we use more textual information with abstracts and a more refined method to extract keywords. \cite{bouveyron2016stochastic} uses both information simultaneously for the classification, capturing thus orthogonal dimensions with more difficulty.

Our contribution is original and significant on at least two aspects :
\begin{enumerate}
	\item we combine endogenous classifications in a network multilayer fashion;
	\item a large dataset is constructed from scratch to study a journal not referenced in main databases, tackling both data retrieval and large scale data processing issues.
\end{enumerate}

\cite{light2014open} already introduced a large scale open database and associated tools for the study of science. Our work is complementary as our tool allow the retrieval of targeted data which may not be referenced in such databases.

The rest of the paper is organized as follows : we describe in the next section the dataset used and the data collection procedure. We then study properties of the citation network and describe the procedure to construct the semantic classification through text-mining. We finally study complementary measures of interdisciplinarity obtained with the different classifications.

\section*{Database Construction}
\label{sec:data}

\begin{figure}
\includegraphics[width=\linewidth]{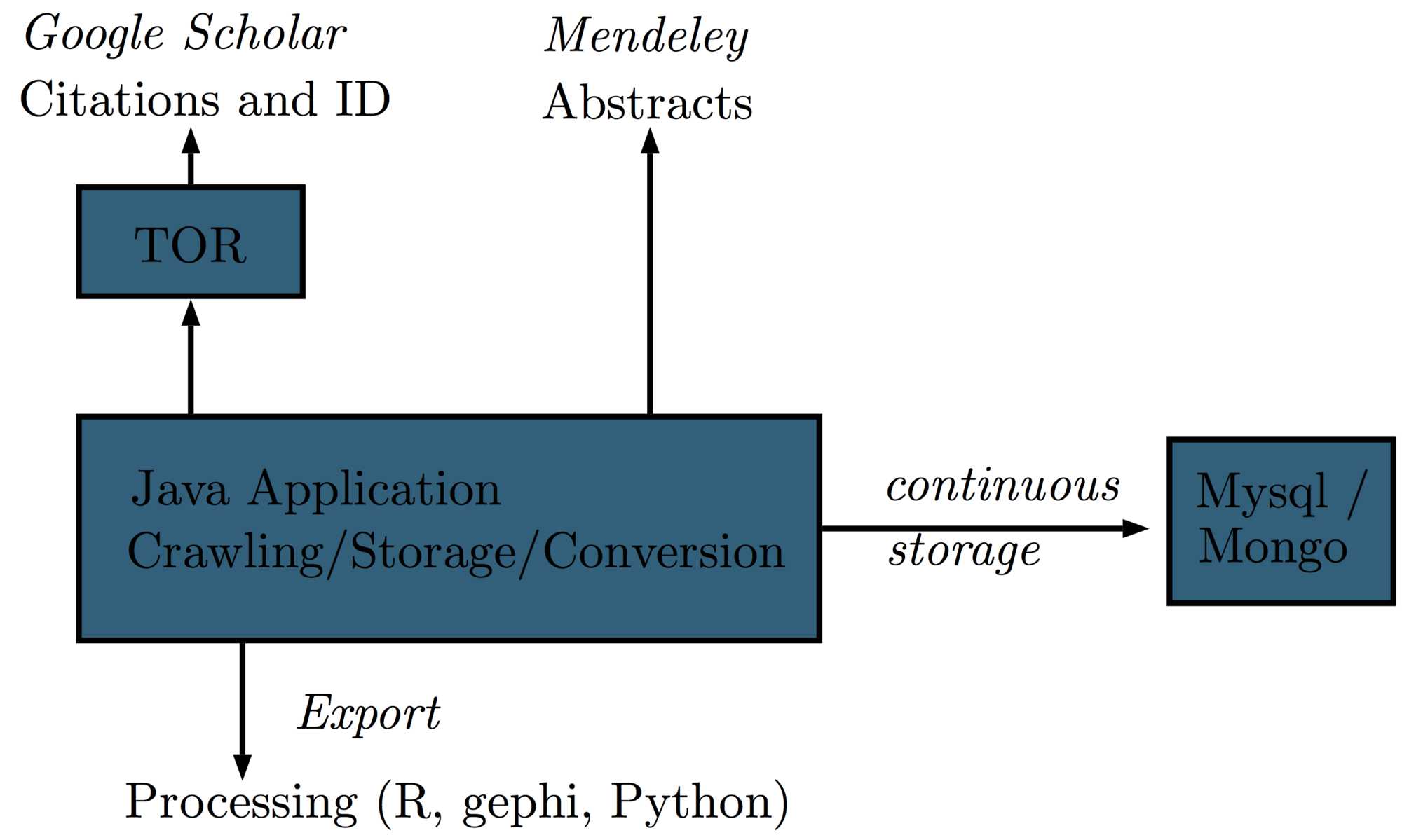}
\caption{\textbf{Heterogeneous Bibliographical Data Collection and processing.} Architecture of the application for content (semantic data), metadata and citation data collection. The heterogeneity of tasks requires the use of multiple languages : data collection and management is done in Java, and data stored in databases (Mysql and MongoDB) ; data processing is done in python for Natural Language Processing and in R for statistical and network analyses; graph visualizations are done with Gephi software.}
\label{fig:datacollection}
\end{figure}

Our approach imposes some requirements on the dataset used, namely: (i) cover a certain neighborhood of the studied journal in the citation network in order to have a consistent view on the scientific landscape; (ii) have at least a textual description for each node. For these to be met, we need to gather and compile data from heterogeneous sources. We use therefore an application specifically designed, which general architecture is given in Fig.~\ref{fig:datacollection}. Source code of the application and all scripts used in this paper are available on the open \texttt{git} repository of the project\footnote{at \texttt{https://github.com/JusteRaimbault/HyperNetwork}}. The data collection application is written in \texttt{Java}, natural language processing in \texttt{python}, and network analyses in \texttt{R} using the \texttt{igraph} package. Network visualisation are done with the \texttt{Gephi} software. Database used are \texttt{MySql} (production base of the journal) and \texttt{MongoDB} (semantic network construction). Raw and processed data are also openly available on Dataverse\footnote{at \texttt{http://dx.doi.org/10.7910/DVN/VU2XKT}}. We recall that an important contribution of this paper is the construction of such an hybrid dataset from heterogeneous sources, and the development of associated tools that can be reused and further developed for similar purposes.

\subsection*{Initial Corpus}

The production database of \textit{Cybergeo} (snapshot taken in February 2016, provided by the editorial board), provides after pre-processing the initial database of articles, with basic information (title, abstract, publication year, authors). The processed version used is available together with the full database constructed, as a \texttt{mysql} dump, at the address given above. This base provide also bibliographical records of articles that give all references cited by the initial base (\emph{forward citations} for the initial corpus).

\subsection*{Citation Data}

Citation data is collected from \texttt{Google Scholar}, that is the only source for incoming citations~\citep{noruzi2005google} in our case as the journal is poorly referenced in other databases\footnote{or was just added as in the case of \textit{Web of Science}, indexing \textit{Cybergeo} since May 2016 only}. We are aware of the possible biaises using this single source (see e.g.~\cite{bohannon2014scientific})\footnote{or \texttt{http://iscpif.fr/blog/2016/02/the-strange-arithmetic-of-google-scholars}}, but these critics are more directed towards search results or possible targeted manipulations than the global structure of the citation network. The automatic collection requires the use of a crawling software to pipe requests, namely \texttt{TorPool}~\citep{torpool} that provides a Java API allowing an easy integration into our application of data collection. A crawler can therethrough retrieve html pages and get backward citation data, i.e. all citing articles for a given initial article. We retrieve that way two sub-corpuses: references citing papers in \textit{Cybergeo} and references \emph{citing the ones cited} by \textit{Cybergeo}. At this stage, the full corpus contains around $4\cdot10^5$ references.

For the sake of simplicity, we will denote by \emph{reference} any standard scientific production that can be cited by another (journal paper, book, book chapter, conference paper, communication, etc.) and contains basic fields (title, abstract, authors, publication year). We work in the following on networks of references, linked by citations.

\subsection*{Text Data}

\begin{figure}
\centering
\includegraphics[width=\linewidth]{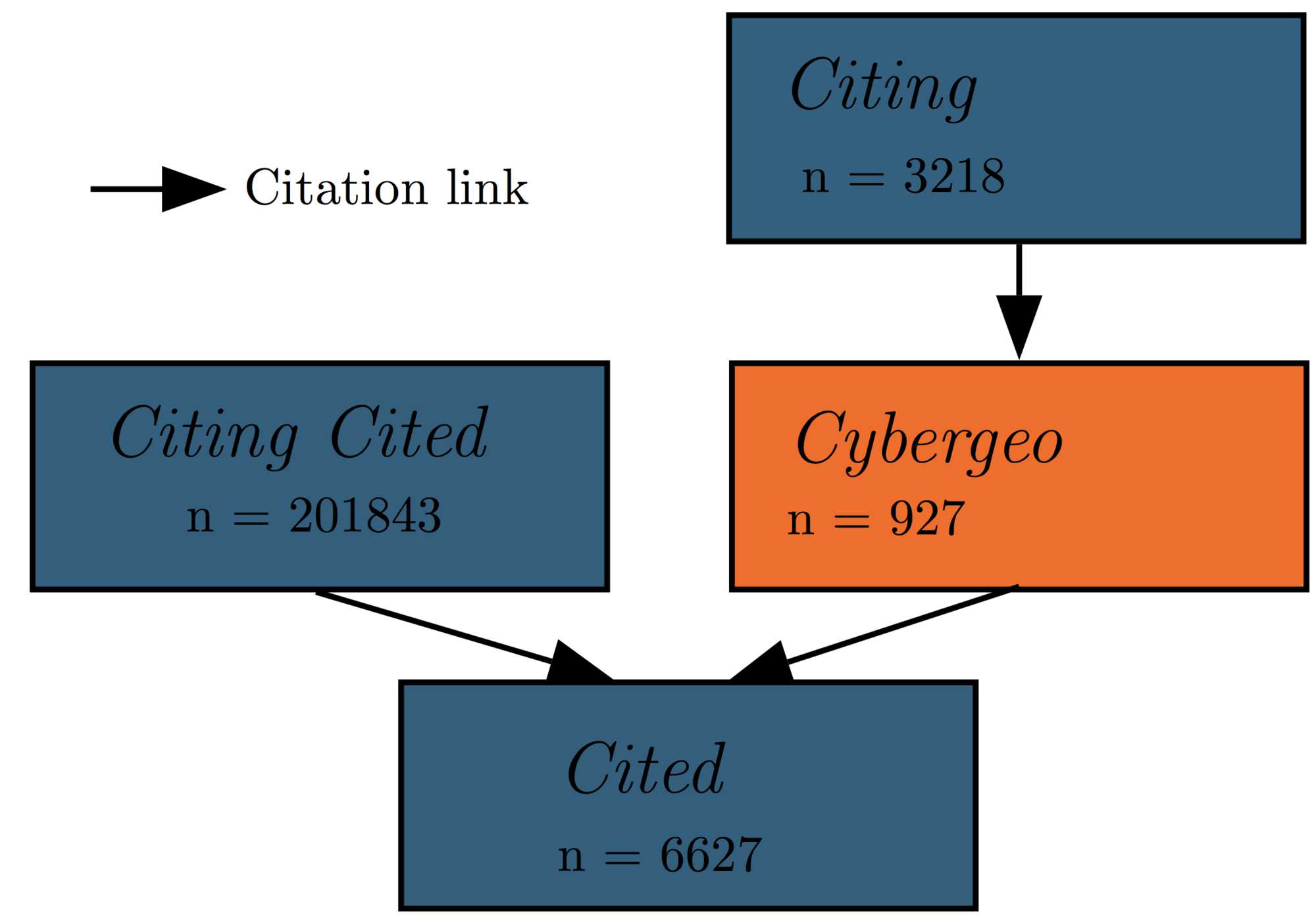}
\caption{\textbf{Structure and content of the citation network.} The original corpus of \emph{Cybergeo} is composed by 927 articles, themselves cited by a slightly larger corpus (yielding a stationary impact factor of around 3.18), cite $\simeq 6600$ references, themselves co-cited by more than $2\cdot 10^5$ works for which we have a textual description.}
\label{fig:citationnetwork}
\end{figure}

A textual description for all references is necessary for a complete semantic analysis. We use for this an other source of data, that is the online catalog of \textit{Mendeley} reference manager software~\cite{mendeley}. It provides a free API allowing to get various records under a structured format. Although not complete, the catalog provides a reasonable coverage in our case, around 55\% of the full citation network. This yields a final corpus with full abstracts of size $2.1\cdot 10^5$. The structure and descriptive statistics of the corresponding citation network is recalled in Fig.~\ref{fig:citationnetwork}.

\section*{Methods and Results}
\label{sec:results}

\subsection*{Citation Network Properties}

\subsubsection*{Properties}

As detailed above, we are able by the reconstruction of the citation network at depth $\pm 1$ from the original $927$ references of the journal to retrieve around $45\cdot 10^6$ references, on which $2.1\cdot 10^5$ have an abstract text allowing semantic analysis. A first glance on citation network properties provides useful insights. Average in-degree (which gives the cumulated number of citations since a reference was published) on references for which it can be defined has a value of $\bar{d}=121.6$, whereas for articles in \textit{Cybergeo} we have $\bar{d}=3.18$. This difference suggests a variety for status of references, from old classical works (the most cited has 1051 incoming citations) to recent less influential works.

\begin{figure}
\centering
\includegraphics[width=\linewidth]{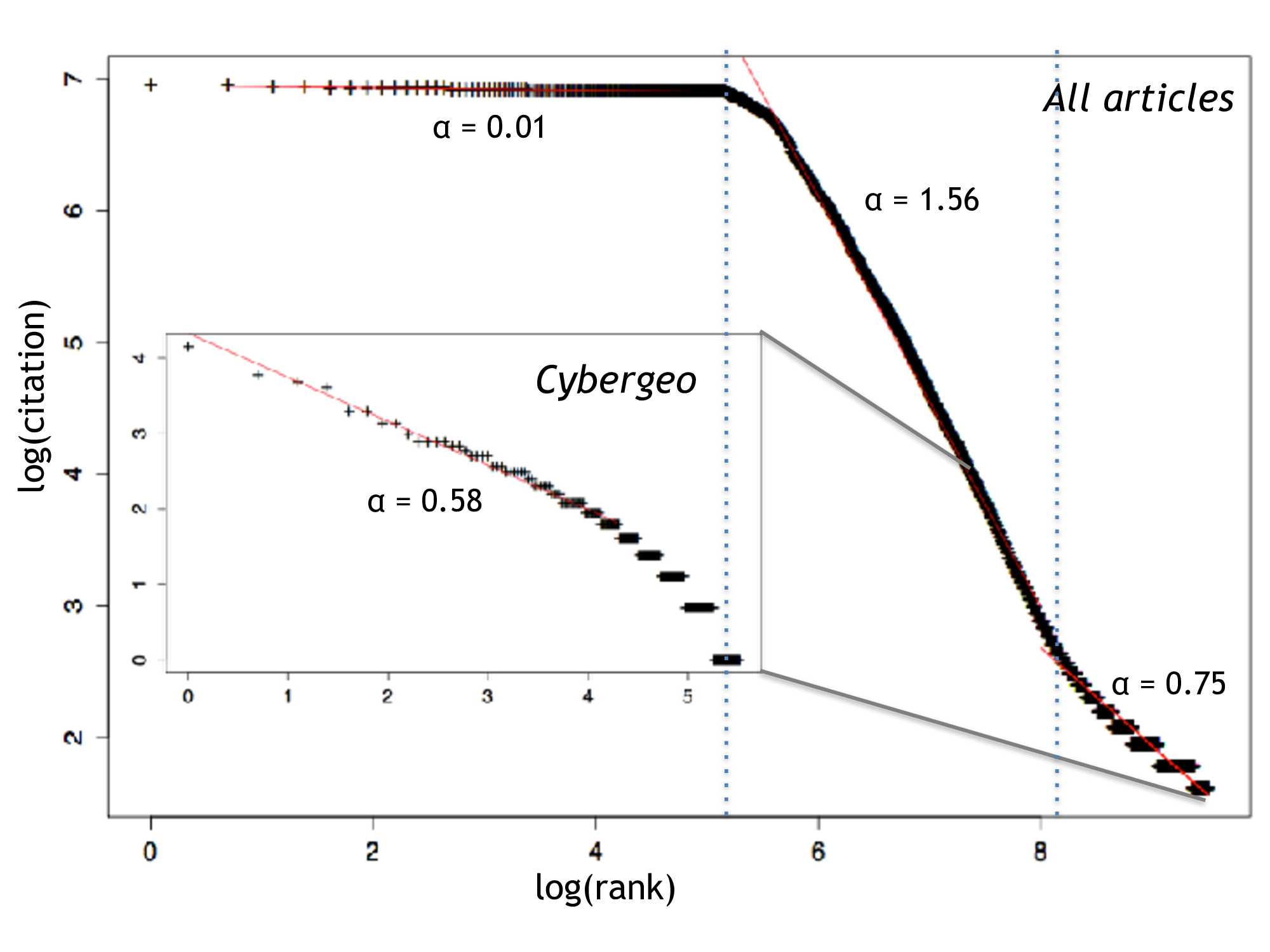}
\caption{\textbf{Rank-size plot of citations received.} The plot unveils three superposed citations regimes, corresponding to power laws with different levels of hierarchy. The references in \textit{Cybergeo} (inset plot) are themselves in the tail and less hierarchical.}
\label{fig:ranksize}
\end{figure}

This diversity is confirmed by the hierarchical organisation examined in Fig.~\ref{fig:ranksize} that unveils three superposed regimes. More precisely, we look at the rank-size plot, given by the logarithm of the number of citations received as a function of the rank of the paper. Scaling properties, which emerge in several models of network growth and are pervasive in real-world networks, are a powerful tool to understand the organisation of complex systems \citep{Barabasi509}.

We find, as expected~\citep{redner1998popular}, localized power-law behaviors. A first set of around 150 references shows a very low hierarchy (rank-size exponent $\alpha = 0.01$) and corresponds to classical references in different disciplines. A second regime ($\alpha = 1.56$) is much more hierarchized, followed by a last regime less hierarchical ($\alpha = 0.75$) containing more recent papers (average publication year mid-2005, against mid-1998 for the second and 1983 for the first).

\begin{figure}
\includegraphics[width=\linewidth]{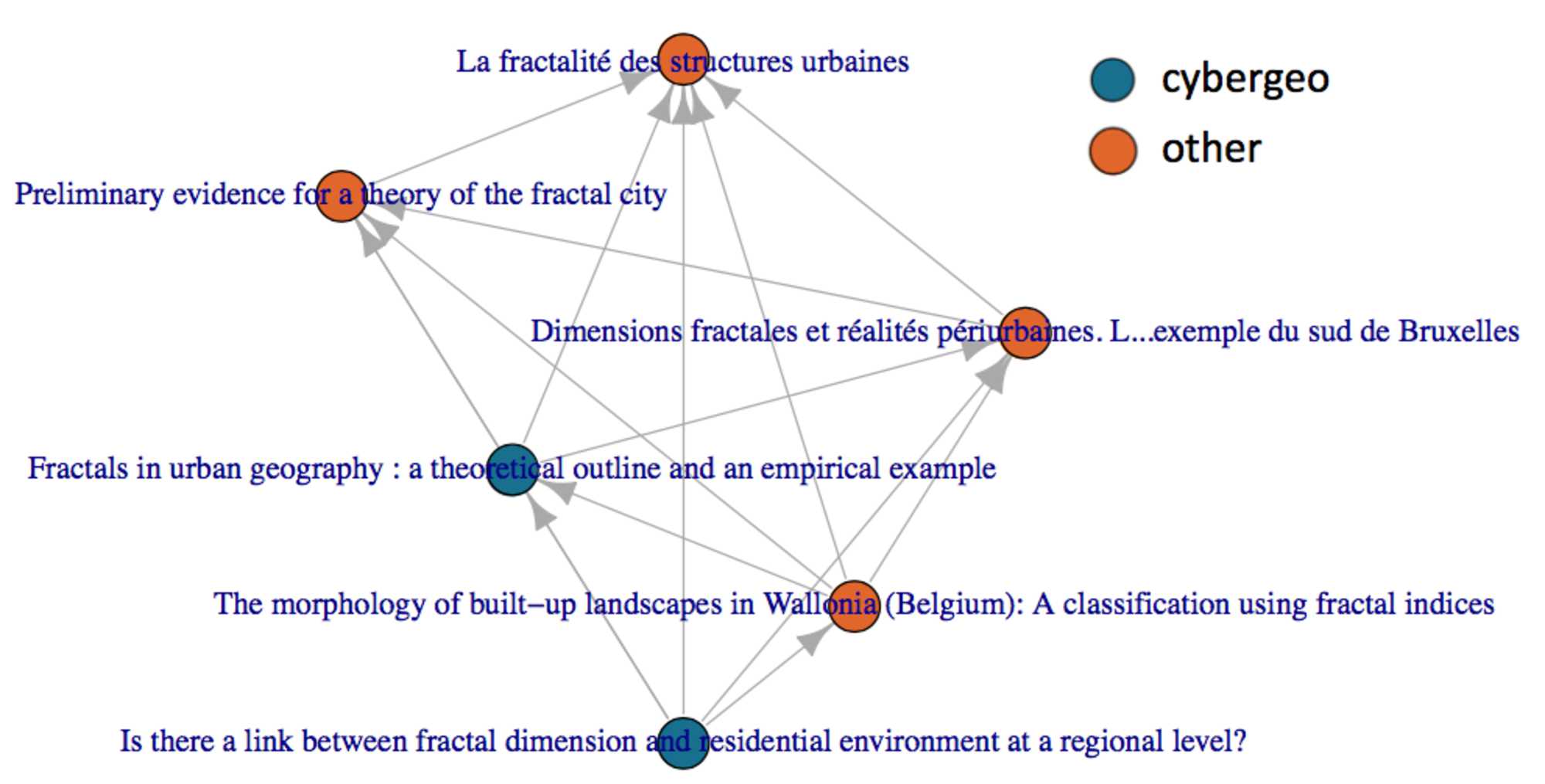}
\caption{Example of a maximal clique in the citation network, paper of \texttt{cybergeo} being in blue. Such topological structure reveal citation practices such as here a systematic citation of previous works in the research niche.}
\label{fig:cliques}
\end{figure}

Other topological properties reveal typical patterns of citation practices, as for example the existence of high-order cliques. Cliques are subset of nodes between which all possible connections exist. Their existence implies citation practices in which all previous papers are systematically cited in new works. The compatibility of this process with the cumulative nature of knowledge may be questionable~\citep{pumain2005cumulativite}, since the previous knowledge production process is reconstructed each time instead on relying only on the most recent state of knowledge. An exemple of such a clique in shown in Fig.~\ref{fig:cliques}, where 6 publications studying the fractal nature of urban structures all cite the previous publications in the clique.

\subsubsection*{Citation communities}

\begin{figure}
\includegraphics[width=\linewidth]{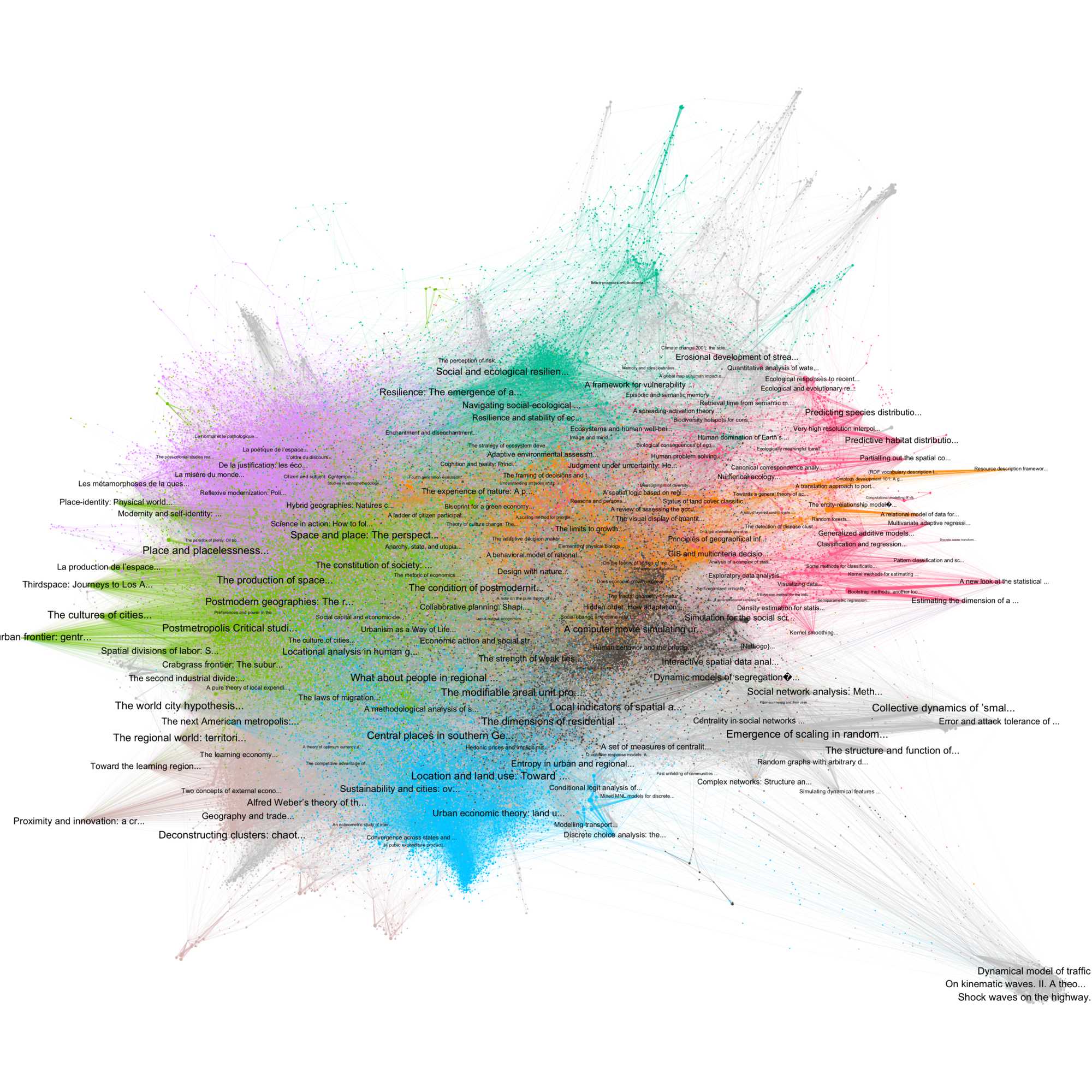}
\caption{\textbf{Citation Network.} The directed citation network is composed by references as nodes and citation links between these. We show only the ``core'' of the citation network, composed by references with a degree larger than one ($\left|V\right| = 107164$ and $\left|E\right| = 309778$). Nodes and edges color gives the main community (for example ecology in magenta, GIS in orange, Socio-ecology in turquoise, Social geography in green, Spatial analysis in blue). Node labels give shortened titles of most cited papers, size is scaled according to their in-degree. The graph is spatialized using a Force-Atlas algorithm.}
\label{fig:citnw}
\end{figure}

The citation network is a first opportunity to construct endogenous disciplines, by extracting citation communities. More precisely, this step aims at finding recurrent patterns in citations that would define a field by its citation practices. In order to be consistent with the particular data structure we have (missing incoming citations for sub-corpuses at maximal depth), we filter the network by removing all nodes with degree smaller than one. This ensures that the nodes kept are either at least cited by an other node (and thus there are no missing edges for these nodes) or cite at least two other nodes, what can make ``bridges'' between sub-communities. The resulting network has a size of $\left|V\right| = 107164$ nodes and $\left|E\right| = 309778$ edges. The citation network is visualized in Fig.~\ref{fig:citnw}.

We use a standard modularity optimization algorithm to identify communities~\citep{blondel2008fast} in the citation network. This Louvain algorithm is applied on the corresponding undirected network following the elementary solution for community detection in directed networks \citep{malliaros2013clustering}. It provides 29 communities with a modularity of 0.71, which means that communities are highly integrated (generally, values above 0.4 are already considered as strongly clustered \citep{newman2006modularity}). In comparison, a bootstrap of 100 randomisations of links in the network gives an average modularity of $-1.0\cdot 10^{-4} \pm 4.4\cdot 10^{-4}$. This bootstrap gives statistical significance to the value we obtained, as the observed modularity is larger by more than four order of magnitude than the standard deviation of this null model.

We name the communities by inspection of the titles of most cited references in each. This naming process was done under the supervision of the editorial head of the journal. The 14 communities that have a size larger than 2.5\% of the network are : Complex Networks, Ecology, Social Geography, Sociology, GIS, Spatial Analysis, Agent-based Modeling and Simulation (ABMS), Socio-ecology, Urban Networks, Urban Simulation, Urban Studies, Economic Geography, Accessibility/Land-use, Time Geography. These categories do not directly correspond to well-defined disciplines, as some correspond more to methods (ABMS), objects of study (Urban Studies), or paradigms (Complex Networks). Some are ``specializations'' of others : most papers in Urban Studies can also be classified as Critical and Social geography. This way, we construct endogenous disciplines that correspond to \emph{scientific practices} (what is cited) more than their representation (the ``official'' disciplines). The relative positioning of communities in Fig.~\ref{fig:citnw}, obtained with a Force-Atlas algorithm, tells a lot about their respective relations : for example, social geography makes a bridge between Urban Studies and Economic Geography, whereas the connection between Socio-ecology and Urban simulations is done by GIS (what can be expected as geomatics is an interdisciplinary field). GIS also separates and connects two subfield of Ecology, on one side more thematic studies on ecological habitats, and on the other sides statistical methods. These relations already inform qualitatively patterns of interdisciplinarity, in the sense of integration measures. We will also in the following use these communities to situate the semantic classification.

Regarding the validity of the map we obtained, the content and structure of our dataset does not allow us to give an exogenous validation using existing techniques \cite{boyack2005mapping}. We rely for the validity of our approach (i) first on the expert knowledge used to name the communities; (ii) secondly on the high values of the modularity which witness a strong endogenous relevance; (iii) the fact that we do not aim to produce exhaustive maps of the discipline, but only to explore the neighborhood of the origin journal; and (iv) the low sensitivity to network perturbations as it is developed below in the sensitivity analysis section.

\subsection*{Semantic Communities Construction}

We now turn to the methodological details for the construction of the semantic classification. This step adapts the methodology described by~\cite{bergeaud2017classifying}, who construct a semantic classification on patent data.

\subsubsection*{Relevant Keywords Extraction}

We recall that our corpus with available text consists of around $2\cdot 10^5$ abstracts of publications at a topological distance shorter than 2 from the journal \textit{Cybergeo} in the citation network. The first important step is to extract relevant keywords from abstracts. Text processing is done with the python library \texttt{nltk}~\citep{bird2006nltk}. We add a particular treatment to the method of~\cite{bergeaud2017classifying}, as our corpus is multilingual: language detection is done with the technique of \emph{stop-words}~\citep{baldwin2010language}. We also use a specific tagger (the function allowing the attribution of grammatical function to words), \texttt{TreeTagger}~\citep{schmid1994probabilistic}, for languages other than English.

To summarize, the keyword extraction workflow goes through the following steps :

\begin{enumerate}
\item Language detection is done using \textit{stop-words}, by selecting the language which \texttt{nltk} stop-words dictionary has the most words in common with the current text.
\item Pos-tagging (detection of word functions) and stemming (extraction of the \emph{stem}) are done differently depending on language :
\begin{itemize}
\item English : \texttt{nltk} built-in pos-tagger, combined to a \emph{PorterStemmer}
\item French or other : use of \texttt{TreeTagger}~\citep{schmid1994probabilistic}
\end{itemize}
\item Selection of potential \textit{n-grams} (keywords of length $n$ with $1 \leq n \leq 4$) following the given grammatical rules to identify noun phrases: for English $\bigcap \{NN \cup VBG \cup JJ \}$ (which correspond respectively to noun, gerund verb, adjective), and for French $\bigcap \{NOM \cup ADJ\}$ (respectively noun, adjective). More elaborated procedures to infer grammatical rules for noun phrases by learning from corpuses do exist \citep{cardie1998error}, but are devised to gain knowledge on grammar specifically. We follow the simple procedure of \cite{chavalarias2013phylomemetic} with these simple fixed rules for noun phrases, similarly to \cite{kumar2008automatic} without the use of heavy additional dictionaries for filtering. Other languages are a negligible proportion of the corpus and are discarded.
\item Estimation of the relevance \textit{n-grams}, by attributing a score following the deviation of the statistical distribution of co-occurrences to a random distribution.
\end{enumerate}

\subsubsection*{Semantic Network}

We keep at this stage a fixed number $K_W$ of \textit{n-grams}, based on their relevance score, that will be designated as the relevant keywords. We find that for large values of $K_W$, results are not sensitive to the total number of keywords, and take a reasonably large value for computational performance, $K_W = 50,000$. We construct the co-occurrence matrix of the relevant keywords. This co-occurrence matrix provides the semantic network as its adjacency matrix : nodes are keywords, and they are linked according to their co-occurrences.

\subsubsection*{Sensitivity Analysis}

We observe the same phenomenon than in~\cite{bergeaud2017classifying}, that is the existence of nodes with large degree and not specific to a particular field : for example \texttt{model} and \texttt{space} are used in most of subfields of Geography. We also adapt the original filtering procedure, as we do not have here an exogenous information to calibrate parameters. We assume the highest degree terms do not carry specific information on particular classes and can be thus filtered given a maximal degree threshold $k_{max}$. We keep the second filter on a minimal edge weight threshold $\theta_w$. We add the supplementary constraint that keywords are also filtered on a document frequency window $\left[ f_{min},f_{max} \right]$ (number of references in which they appear), what is slightly different from network filtering.

\begin{figure}
\centering
\includegraphics[width=\linewidth]{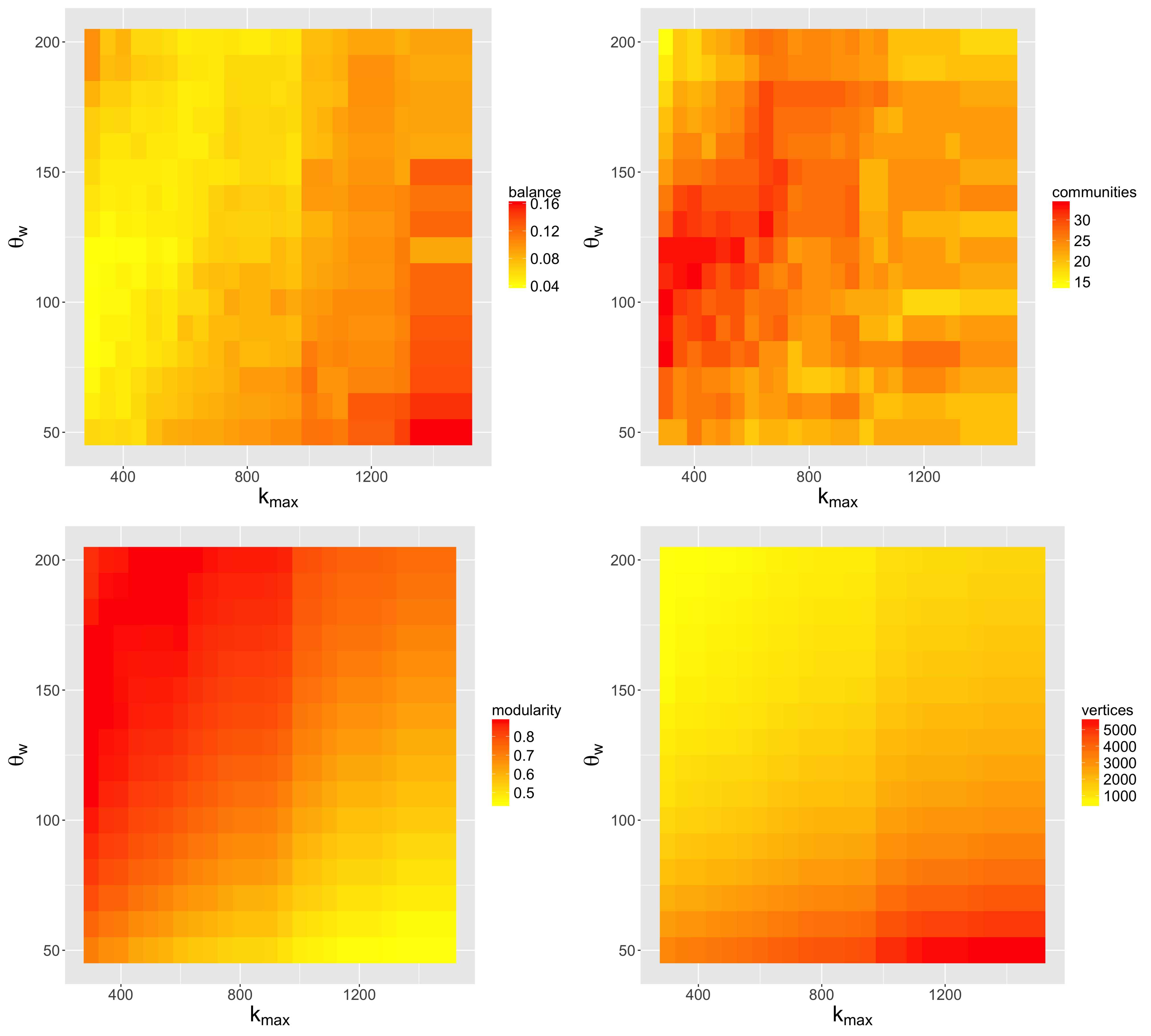}
\caption{\textbf{Sensitivity analysis of network indicators to filtering parameters.} We show here 4 indicators (balance between community sizes, modularity of the decomposition, number of communities, number of vertices), as a function of parameters $k_{max}$ and $\theta_w$, at fixed $f_{min} = 50, f_{max} = 10000$. Close values for these two last parameters (in a reasonable range) give similar behavior.}
\label{fig:sensitivity}
\end{figure}

A sensitivity analysis of resulting network topology to these four parameters is presented in Fig.~\ref{fig:sensitivity}. Given a filtered network, we detect communities using modularity optimization as before for the citation network. Various properties of the network can be optimized, and we look in particular at its size (number of keywords after filtering), the optimal modularity, the number of communities, and the balance between their sizes (defined as a concentration index $\sum_k s_k^2 / (\sum_k s_k)^2$). This multi-objective optimization problem does not have a unique solution as objectives are contradictory in a complex way, and a compromise point must be chosen. Indeed, one can obtain a very high modularity but with a small network which will finally cover only a small fraction of the corpus, whereas a large network yields lower modularity values as shown in Fig.\ref{fig:sensitivity}. The process is similar for other indicators.

We take a compromise point between modularity and network size, with a high balance and a reasonable number of communities, given by $k_{max} = 1200, \theta_w = 100, f_{min} = 50, f_{max} = 10000$. This point was taken on the compromise line between modularity and network size (roughly around the line linking $(\theta_w=50,k_{max}=400)$ with $(\theta_w=100,k_{max}=1200)$), with a high balance (taking $k_{max}\geq 1000$) and not too much communities. These values give a network of size 2868, with 18 communities and a modularity of 0.57.

Note that the small proportion of keywords in French is always separated from the rest of the network as they cannot co-occur with English keywords, and that with these parameter settings no French keywords are kept. All communities described in the following therefore contain only keywords in English.

\subsubsection*{Semantic Communities}

We obtain therein communities in the semantic network with the optimized filtering parameters. At the exception of a small proportion apparently resulting from noise (representing less than 10 keywords in 3 communities that we remove, i.e. 0.33\% of keywords), communities correspond to well-defined scientific fields, domains, or approaches. Naming is also done by inspection of the most relevant keywords in each community, in order to stick here to a certain level of supervision.

\begin{table}
\caption{\textbf{Semantic communities reconstructed from community detection in the semantic network.}}
\label{tab:domains}
\hspace{-2cm}
\begin{tabular}{lll}
\hline\noalign{\smallskip}
Name & Size & Keywords  \\
\noalign{\smallskip}\hline\noalign{\smallskip}
Political sciences/critical geography & 535 & \texttt{decision-mak, polit ideolog, democraci, stakehold, neoliber} \\
Biogeography & 394 & \texttt{plant densiti, wood, wetland, riparian veget} \\
Economic geography & 343 &  \texttt{popul growth, transact cost, socio-econom, household incom} \\
Environnment/climate & 309 & \texttt{ice sheet, stratospher, air pollut, climat model} \\
Complex systems & 283 & \texttt{scale-fre, multifract, agent-bas model, self-organ} \\
Physical geography & 203 & \texttt{sedimentari, digit elev model, geolog, river delta} \\
Spatial analysis & 175 & \texttt{spatial analysi, princip compon analysi, heteroscedast, factor analysi} \\
Microbiology & 118 & \texttt{chromosom, phylogenet, borrelia} \\
Statistical methods & 88 & \texttt{logist regress, classifi, kalman filter, sampl size} \\
Cognitive sciences & 81 & \texttt{semant memori, retrospect, neuroimag} \\
GIS & 75 & \texttt{geograph inform scienc, softwar design, volunt geograph inform, spatial decis support} \\
Traffic modeling & 63 & \texttt{simul model, lane chang, traffic flow, crowd behavior} \\
Health & 52 & \texttt{epidem, vaccin strategi, acut respiratori syndrom, hospit} \\
Remote sensing & 48 & \texttt{land-cov, landsat imag, lulc} \\
Crime & 17 & \texttt{crimin justic system, social disorgan, crime} \\
\noalign{\smallskip}\hline
\end{tabular}
\end{table}

\begin{figure}
\includegraphics[width=\linewidth]{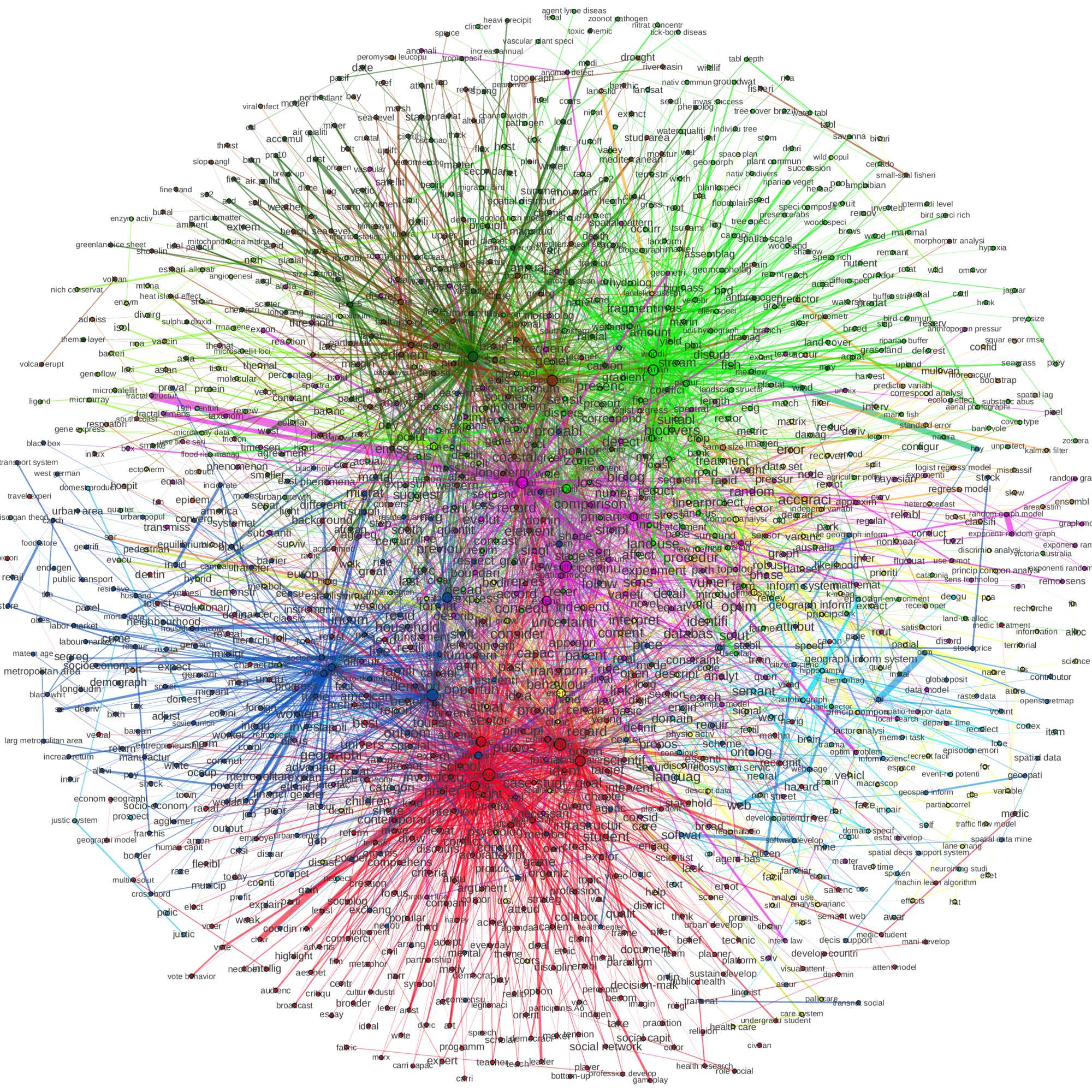}
\caption{\textbf{Visualization of the semantic network.} Network is constructed by co-occurrences of most relevant keywords. Filtering parameters are here taken according to the multi-objective optimization done in Fig.~\ref{fig:sensitivity}, i.e. $(k_{max}=1200,\theta_w=100,f_{min}=50,f_{max}=10000)$. The graph spatialization algorithm (Fruchterman-Reingold), despite its stochastic and path-dependent character, unveils information on the relative positioning of communities.}
\label{fig:semanticnw}
\end{figure}

\begin{figure}
\centering
\includegraphics[width=\linewidth]{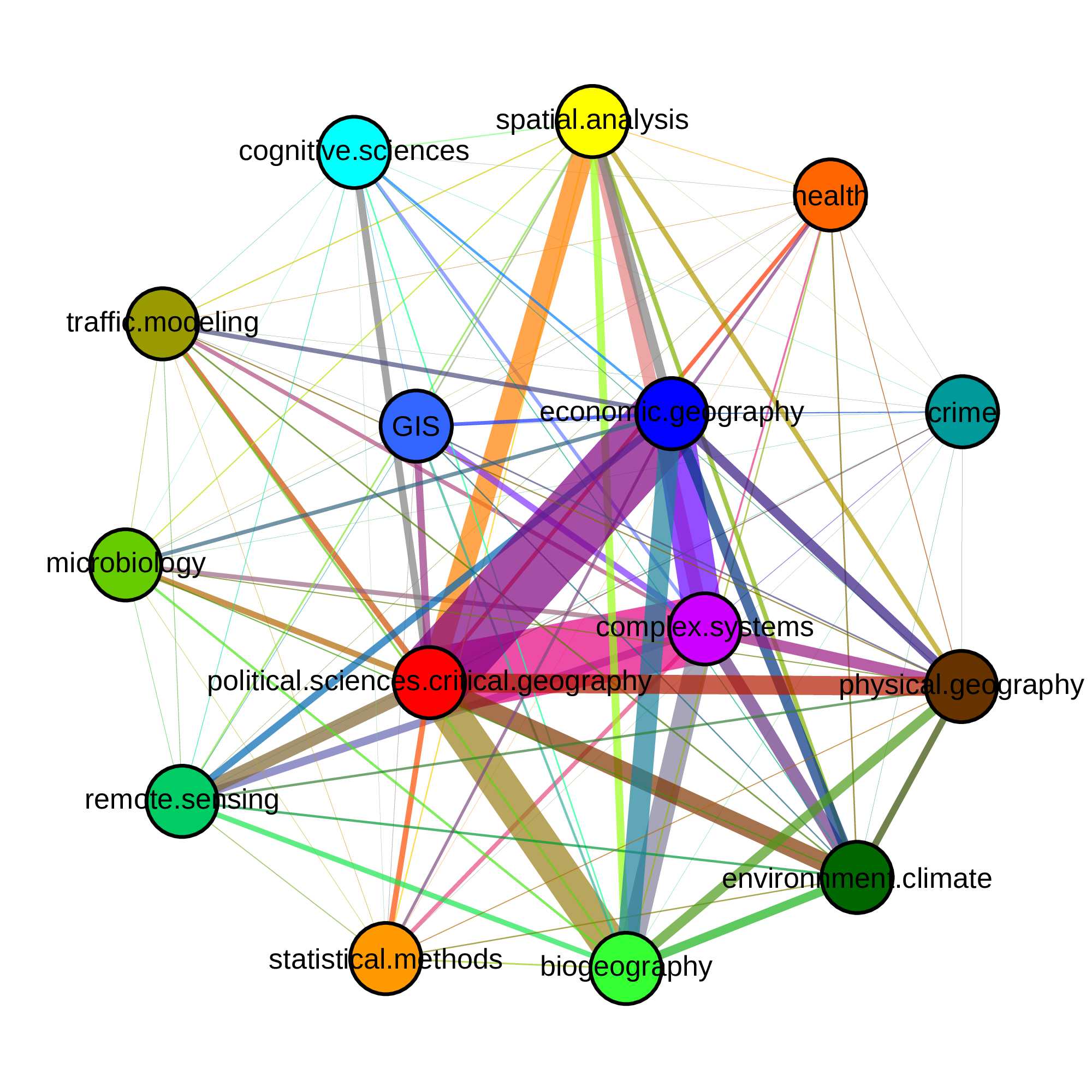}
\caption{\textbf{Synthesis of semantic communities and their links.} Weights of links are computed as probabilities of co-occurrences of corresponding keywords within references. Once these weights have been calculated with the composition of references in terms of semantic probabilities (see below), the graph is spatialized using a Fruchterman-Reingold algorithm with the Gephi software.}
\label{fig:comsynthesis}
\end{figure}

Table~\ref{tab:domains} summarizes the communities, giving their names, sizes, and corresponding keywords. The most important community is related to issues in political science and critical geography, what could have been expected as several previously obtained citations communities (Social geography, Urban studies) deal with these issues. We then obtain a large cluster of terms related to biogeography, that must correspond to publications in Ecology and Socio-ecology identified before, together with a community in Environment and Climate.

In a way similar to the citation communities, but more pronounced here, we obtain endogenous ``disciplines'' that can correspond to real disciplines, to methodologies, to object of studies. This classification thus also unveil \emph{effective scientific practices}, here in terms of semantic content. A class here related to complex systems can be associated to a paradigm and various approaches that were separated in the citation communities : agent-based models and complex networks for example. On the contrary, some studies that were gathered in a large domain before can be precisely differentiated in the semantic network, such as microbiology and health here that are used by studies related to socio-ecology or ecology in the citation network. Some very specific domains appear here as they have very few connections in their actual semantic content : for example, Geography of crime is very precise and disconnected from other communities.

We show in Fig.~\ref{fig:semanticnw} a visualisation of the semantic network, in which the positioning of communities, induced by a Fruchterman-Reingold algorithm (that we use here to have a more precise layout in the relative positioning compared to Force Atlas~\citep{jacomy2014forceatlas2}). The bridging between distant disciplines is done quite differently compared to the citation network, and reveals thus qualitatively an other dimension of interdisciplinarity, i.e. the semantics shared by disciplines. Here, the communities corresponding to Economic Geography (blue) and to Critical Geography (red) are close as in the citation network, but are linked to ecology and geomorphology (green and brown) by Complex Systems (magenta), although these were not present as a community in the citation network. Complexity methodologies such as Fractals, Scaling~\citep{west2017scale} or Networks~\citep{newman2003structure} are indeed widely used both in social sciences and in physics or biology. The semantic analysis reveals thus that very distant disciplines, that are distant in their citation patterns, are finally close in terms of actual content. This conclusion can naturally subject to the bias that similar terms can correspond to very different ontologies in each disciplines. This actual meaning is partly captured in the embedding of the term in the semantic network, and how it relates to other terms in the semantic network. To what extent ontological proximities can be associated to semantic proximities or reconstructed from the network structure is an open question out of the scope of this work. Terms must thus not be dissociated from their context, which in our case is given by the citation network, thus the interest of the analysis below on the relation between classifications.

In terms of overlaps between communities, in the sense of co-occurrences of corresponding keywords within texts of references, we show a synthesis of links between semantic communities in Fig.~\ref{fig:comsynthesis}. We see that communities such as Critical Geography and Biogeography are not totally disconnected and share still a certain number of co-occurrences. More isolated communities can be spotted such as Health and Crime Geographies. Surprisingly, Statistical Methods does not share strong links with other communities, what could mean that articles dealing with methodological issues in this field are rather disconnected from the field of application, or at least do not describe it extensively. On the contrary, methods in Complex Systems are organically integrated with the thematic issues they tackle.

\subsection*{Sensitivity to corpus definition}

To give a stronger confidence in the results we obtain, we need to study the influence of corpus choice, since (i) our case study is a narrow part of the related disciplines; and (ii) the interaction between missing data and the methodology introduced may play a role in the result obtained, that then could be solely due to spurious effects between both. However, our data collection process and classification method were conceived to be used together: using both layers increase the information that can be extracted, since the collected data has less information than classical datasets.

Our positioning regarding data, i.e. working only with the open dataset in a case of a low data availability, does not allow us to proceed to ``ground truth'' validations with an exogenous classification. We can however study endogenously the sensitivity of the community structure to the corpus definition, by studying the sensitivity to data removal. This controls the process used by data quality and contributes to the validation of the method, rather independently to the dataset used.

\begin{figure}
\centering
\includegraphics[width=\linewidth]{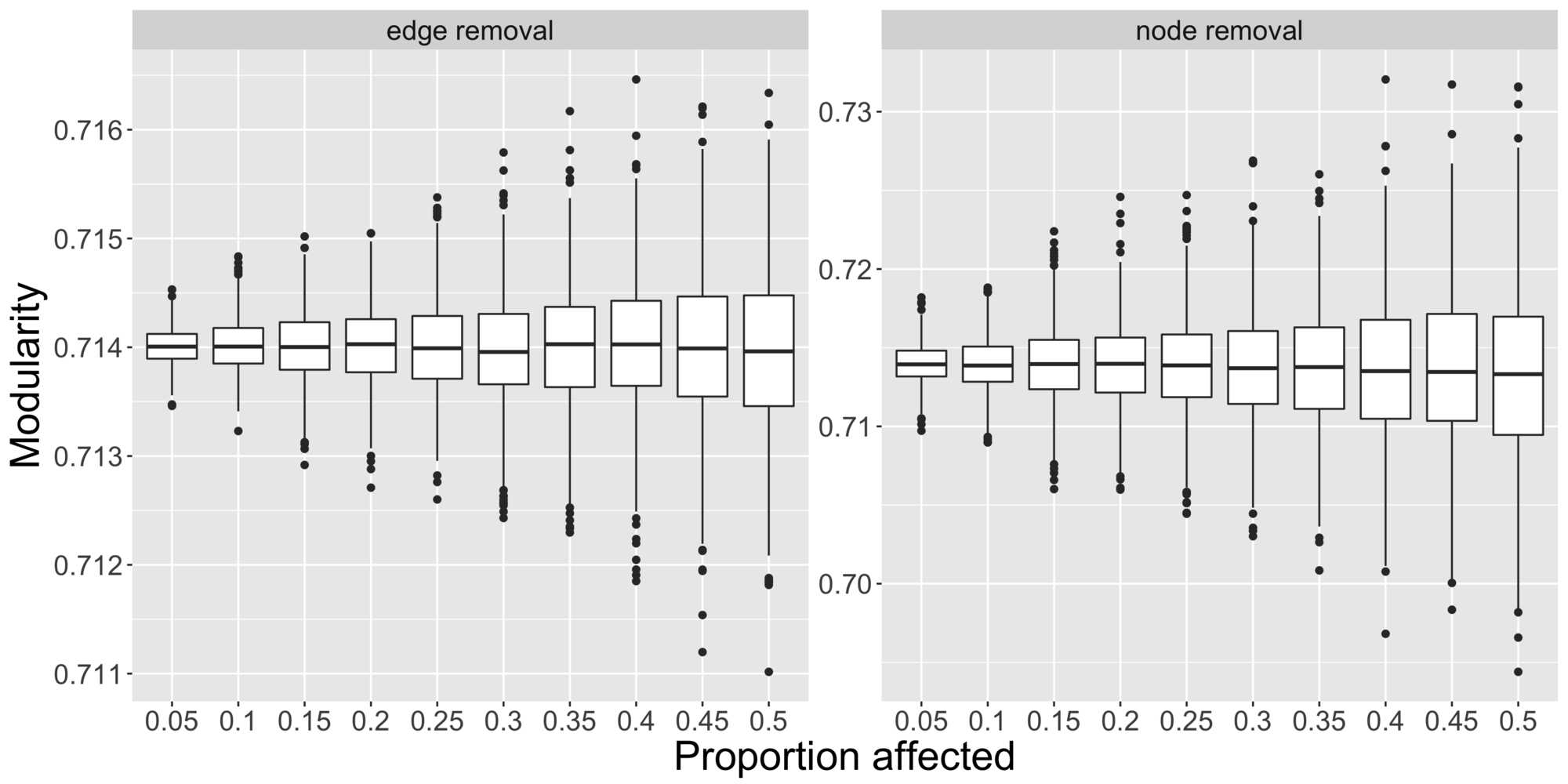}
\caption{\textbf{Sensitivity of the modularity to network modifications.} We give the statistical distribution of modularity as boxplots, as a function of the proportion of entities affected $r$, for edge removal (left) and node removal (right).}
\label{fig:modsens}
\end{figure}

The procedure is the following, according to classical approaches to the study of network robustness \citep{trajanovski2013robustness}: working on the citation network, we test the sensitivity of the community structure to entity removal. A fixed proportion $r$ of nodes or edges are removed randomly, and modularity is estimated in the resulting network. This procedure is bootstrapped $N_b = 1000$ times for each value. Results are shown in Fig.~\ref{fig:modsens}, for $0.05 \leq r \leq 0.5$. We obtain similar qualitative patterns for node and edge removal regarding the form of distributions, but the median modularity slightly decreases for nodes whereas it oscillates for edges. Extreme values are higher for nodes, up to 0.02 difference in absolute value. This remains a very low absolute value, and we find that the community structure has a low sensitivity to data removal, in terms of missing reference (node) or citation link (edge), when going up to 50\% of missing data. This gives confidence in the fact that our method is stable relatively independently of the choice of the origin corpus.

We also tested the influence of the time window used, since our origin corpus spans between 1996 and 2016. We filter the citation network on fiver-year sliding windows covering this span. The size of corresponding subnetworks is relatively stable ($\left| V\right| \in \left[ 27866 ; 60474 \right]$ and $\left| E\right| \in \left[ 18839 ; 33892 \right]$) and the modularities are higher than the full network (between 0.84 and 0.88) and stable. This suggests that the community structure is stable in time.

\subsection*{Semantic composition of citation communities}

\begin{figure}
\centering
\includegraphics[width=\linewidth]{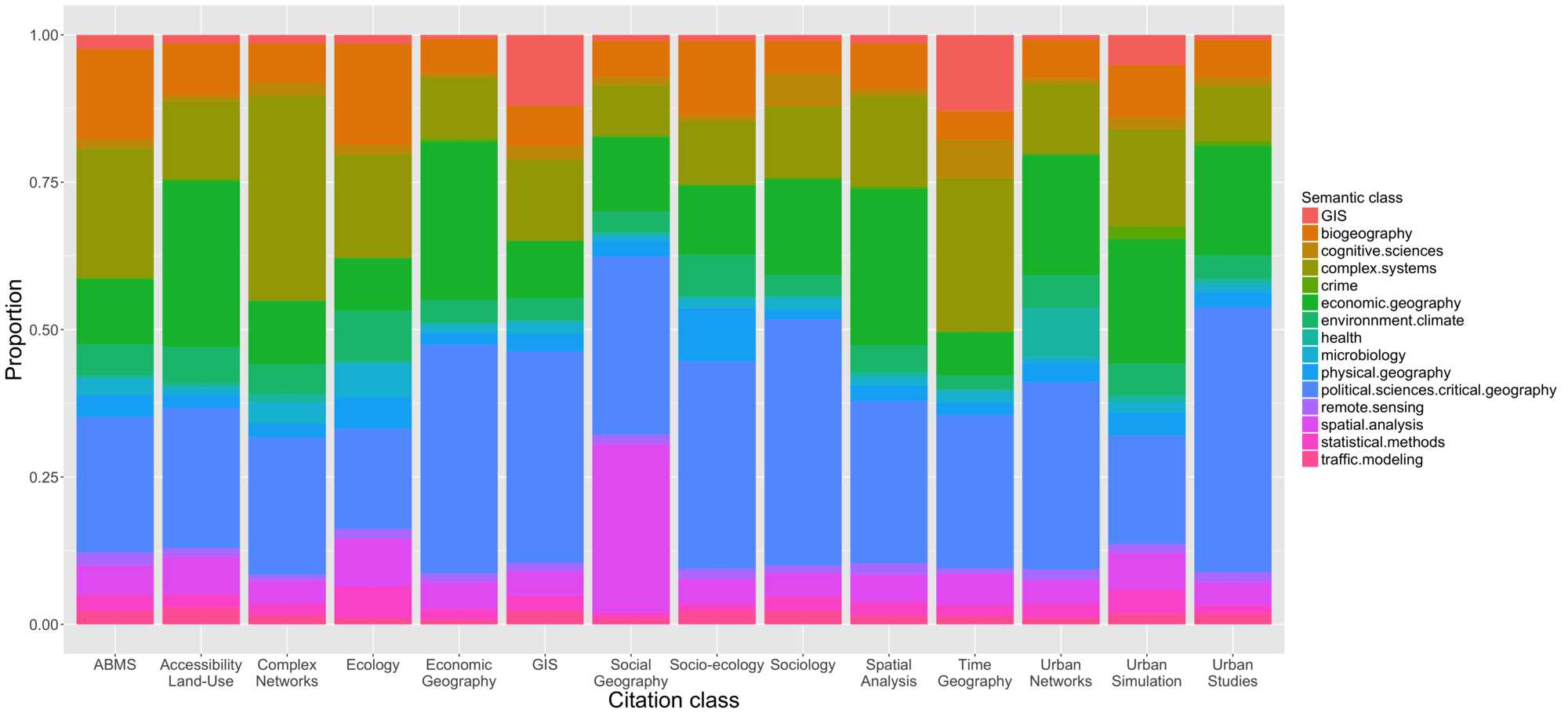}
\caption{\textbf{Composition of citation communities in terms of semantic content.} For each citation class (horizontally), the bar is decomposed as the proportions of each semantic class (given by color).}
\label{fig:citationcontent}
\end{figure}

We can now turn to the study of the relation between classifications. First, a simple way to link them is to look at the semantic content of citation communities. Each reference has a given proportion of keywords within each semantic class, and an average composition in terms of semantic classes for each citation class can thus be computed. We show these composition in Fig.~\ref{fig:citationcontent}. Some expected results are obtained, such as Complex Networks (citation) having the largest part in Complex Systems (semantic), or GIS (citation) the largest in GIS (semantic), and similarly for Economic Geography.

But the study of patterns that could have not been expected is very informative, and unveils practices of interdisciplinarity. For example, Time Geography (citation) uses as much GIS (semantic) as GIS (citation), what means that they should be using the corresponding methods and tools to study the thematic question of spatio-temporal trajectories of geographical agents. The most important in terms of political science (semantic) are Urban Studies, what suggest a convergence of the City as an object of study and of the disciplines of Political Science and Critical Geography. Also interestingly, the citation communities using most biogeography are Ecology (what could have been expected) and ABMS, confirming again the role of the thematic application in complex systems methodologies.

\subsection*{Measuring interdisciplinarity}
\label{subsec:interdisc}

\begin{figure}
\centering
\includegraphics[width=\linewidth]{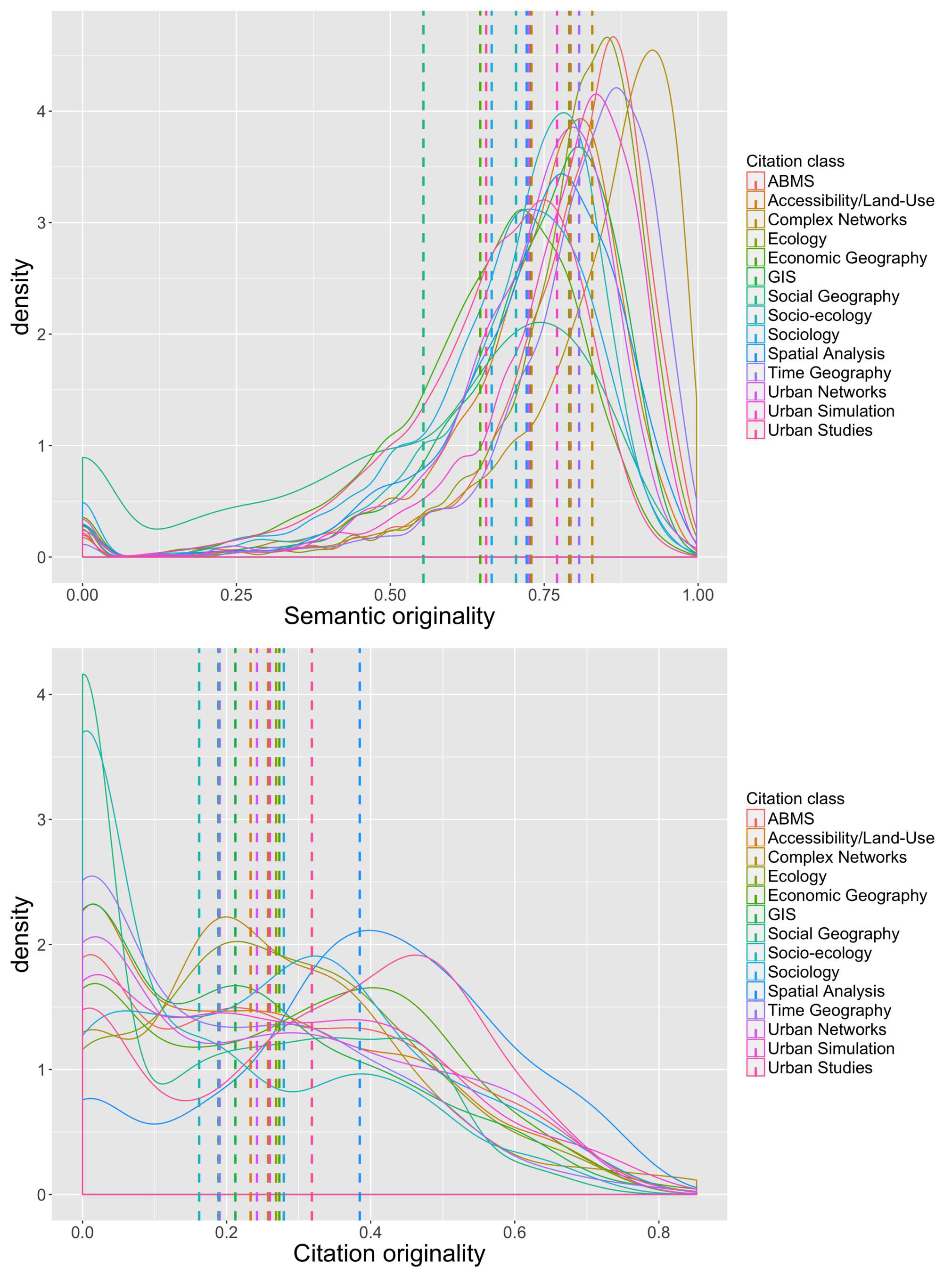}
\caption{\textbf{Statistical distribution of originalities.} We show the smoothed probability densities of originality indexes, by citation class (given by color), for the Semantic originality $o^{(Semantic)}$ (top plot) and for the Citation originality $o^{(Citation)}$ (bottom plot). Dashed lines give the mean for each distribution, with the corresponding color.}
\label{fig:firstorderint}
\end{figure}

We had up to now a qualitative view on interdisciplinarity patterns, by looking at the relative localisation of communities within the citation and semantic classifications, and the relation between the classifications. We propose now to look at quantitative measures of interdisciplinarity, for each classification.

More precisely, for a given classification $C \in \{ Citation,Semantic\}$ a reference $i$ can be viewed as a probability vector $(p_{ij}^{(C)})_j$ on classes $j$ that give for each class the probability to belong to it. Given this setting, we measure interdisciplinarity of one reference using Herfindhal concentration index~\citep{porter2009science}, that can also be called an originality index. We define originality as
\begin{equation}
o_i^{(C)} = 1 - \sum_j {p_{ij}^{(C)}}^2
\label{eq:interdisc}
\end{equation}

For the semantic classification, probabilities are defined as the proportion of keywords of the abstract within each semantic class. With the deterministic citation classification, each reference has only one class and the originality index is always 0. Therefore in order to be able to compare the two classification, we associate a probability to each citation class for each article as the proportion of citations received from this class. The index introduced is original, and measures interdisciplinarity as \emph{how a reference is used} by different disciplines in its lifetime. Indeed, Eq.~\ref{eq:interdisc} yields higher values for more evenly distributed probabilities. When used with semantic probabilities, it translates the originality in terms of language used, whereas with the citation probabilities, it gives the originality in terms of domains using the reference.

We show in Fig.~\ref{fig:firstorderint} the statistical distribution for both indexes $o^{(Semantic)}$ and $o^{(Citation)}$, stratified by citation class. This allow a direct comparison between the two and also an indirect comparison by the variation of semantic distribution between citation classes. For the distribution of semantic originalities, all citation classes exhibit a similar pattern, that is a peak around large values and a smaller peak at zero. It means that either references are highly specialized and have keywords in one class only, or they use keywords from different classes in a quite even manner (for comparison, an abstract with half keywords in a class and half in an other gives an originality of 0.5). The most original, i.e. the most mixed, citation class, is Complex Networks, with a distribution clearly detached from others, what would confirm their use as a method with a lot of different problems. Social Geography is from far the less original, with a large number of single class references, and an average far lower than other classes, what would mean an increased presence of compartmentalization within the associated disciplines.

In terms of citation originality index, the global picture is fundamentally different, as average originality indexes are all lower than 0.4 and most of distributions show their mode in 0, meaning that most references are only cited by their own citation class. Again, Social Geography is the less original, confirming a similar behavior in terms of citation practice than in terms of research content. The most original classes in average, with a peak in large values, are Spatial Analysis and Urban Simulation: this corresponds to the fact that these class feature quite generic methods that can be applied in several fields and are cited accordingly. Complex Networks do not reach the same level, but however exhibit a peak around 0.2 and no peak in 0, together with Ecology, suggesting disciplines having still significant impact in other disciplines.

To summarize, we show (i) different patterns of interdisciplinarity, depending on disciplines, in terms of scientific content (semantic) and of scientific impact (citation); and (ii) a strong qualitative difference in behavior of originalities between the two classifications, what suggests their complementarity.

\subsection*{Correlation between classifications}

In order to strengthen the idea of a complementarity of classifications, that would each capture different dimensions of processes of knowledge production, we finally look at the correlation matrix between classifications. We use this time effective class probabilities for the citation classification, i.e. a vector of zeros expect with a one at the index of the class of the reference. We compute a Pearson correlation coefficient between classes $k$ (in semantic) and $k'$ (in citation) as
\begin{equation}
\rho_{k,k'} = \frac{\Cov \left[ (p^{(Sem)}_{ik})_i , (p^{(Cit)}_{ik'})_i \right]}{\sqrt{\Var\left[(p^{(Sem)}_{ik})_i\right]\Var\left[(p^{(Sem)}_{ik})_i\right]}}
\end{equation}

\noindent where the covariance is estimated with the unbiased estimator.

The structure of the correlation matrix recalls the conclusions obtained when studying the semantic composition of citation communities, such as GIS having a relatively high correlation with GIS ($\rho=0.26$), or Sociology with Political Science ($\rho=0.16$). These correlations are high regarding other values in the matrix, but remain low values in absolute.

More important for our question are summary statistics of the overall matrix. It has a minimum of $-0.16$ (Ecology (citation) against Political Sciences (semantic)), an average of $-0.002$ and a maximum of $0.33$ (Social geography (citation) and Spatial Analysis (semantic)). The ``high'' values are highly skewed, as the first decile is at $-0.06$ and the last at $0.09$, what means that 80\% of coefficient lie within that interval, corresponding to low correlations. In a nutshell, classifications are consistent as highest correlations are observed where one can expect them, but most of classes are uncorrelated, meaning that the classifications are quite orthogonal and therefore complementary.

\section*{Discussion}
\label{sec:discussion}

We have this way shown the complementarity of classifications in the qualitative patterns they unveil, but also quantitatively in terms of interdisciplinarity measures and quantitatively in terms of correlations. Our work can be extended regarding several aspects, of which we give some suggestions below.

\subsection*{Developments}

A first development consists in the comparison of journals. The starting point for construction of the scientific environment, the journal \textit{Cybergeo}, was the entry point but not the subject of our study. A development more focused on journals, trying for example to answer comparative issues, or to classify journals according to their effective level of interdisciplinarity regarding different dimensions, would be potentially interesting. The collection of precise data on the origin of references is however a first step that need to be solved first.

The performance of the semantic classification was also not quantified here. A further validation of the relevance of using complementary information contained in both classifications could be done by the analysis of modularities within the citation network, as done in~\cite{bergeaud2017classifying}. This would however require a baseline classification to compare with, which is not available in the type of data we use. Open repository such as arXiv (for physics mainly) or Repec (for Economics) provide API to access metadata including abstracts, and could be starting points for such targeted case studies.

An other aspect on which our work could shed an interesting light is the univocity of scientific keywords between disciplines. Indeed, the same word can relate to different concepts in different disciplines. Co-occurrences patterns of specific words within the different citation communities should give information on the underlying concepts in each. For example, in the fields we studied, the word ``model'' will correspond to totally different concepts in Quantitative Geography and in Urbanism.

\subsection*{Applications}

A first potential application of our methodology relies on the facts that both classifications unveils thematic domains (objects of study), classical disciplines, methodological communities. These different types of communities can indeed be understood as different \emph{Knowledge Domains}. \cite{raimbault2017applied} postulates co-evolving Knowledge Domains in every process of scientific knowledge production, that are Theoretical, Empirical, Modeling, Methodology, Tools and Data domains. Most of them are necessary for any process, and investigations within one conditions the advances in others. A refinement of classifications, associated with supervised classification to associate knowledge domains to some communities (potentially using full texts to have more precise information on the proportion of each knowledge domains involved in each), would allow to quantify relations between domains. Furthermore, using temporal data with the date of publications, would yield an effective quantification of the \emph{co-evolution} of domains in the sense of patterns of temporal correlations (e.g. Granger causality).

Our work furthermore suggests that the different types of communities each unveil a different structure of science. This has implications for the construction of science maps, which refinement may go towards different maps for each dimension (which can more or less be associated to Knowledge Domains). \cite{wen2017mapping} has for example introduced such specific complementary maps as a new method of ``bibliometric triangulation'' in the case of water research.

An other interesting direction is the application of our classifications to the quantification of spatial diffusion of knowledge, as \cite{maisonobe2013diffusion} does for the diffusion of a specific question in molecular biology. It is not clear if different dimensions of knowledge diffuse the same way: for example citation practices can be correlated to social networks and thus exhibit different patterns than effective research contents. This phenomenon is suggested in terms of a co-evolution for semantics and social networks by \cite{roth2010social}. Therefore, our work would allow to study such questions from complementary point of views.

\subsection*{Positioning}

We finally discuss the positioning of our paper. The analysis developed did not pretend to construct maps of a given discipline, but indeed to allow authors and readers of the journal to better situate their work in their scientific neighborhood. As our approach is fully reproducible and does not require access to restricted databases, we argue that our contribution could foster open science and reflexivity.

We therefore believe the tool we developed can contribute to an increased empowerment of authors and to the development of open science practices. Among the various visions of Open Science~\citep{fecher2014open}, the opening of data is always an important aspect, together with a development of reflexivity in all disciplines, beyond the sole Social Sciences to which it is classically associated. The first point is dealt with by our open tools for dataset construction, whereas the second is implied by the  knowledge of the different dimensions of the scientific environment we studied, allowed by the multilayer methodology introduced. Using other dimensions and methodologies, \cite{banos2018spatialised} illustrates this approach through an interactive application to analyse the corpus of the Cybergeo journal.

\section*{Conclusion}
\label{sec:discussion}

We have introduced a multi-dimensional approach to the understanding of interdisciplinarity, based on citation network and semantic network analysis. Starting from a generalist journal in Geography, we construct a large corpus of the citation neighborhood, from which we extract relevant keywords to elaborate a semantic classification. We then show qualitatively and quantitatively the complementarity of classifications. The methodology and associated tools are open and can be reused in similar studies for which data is difficult to access or poorly referenced in classical databases.

\section*{Acknowledgements}

The author would like to thank the editorial board of Cybergeo, and more particularly Denise Pumain and Christine Kosmopoulos, for having offered the opportunity to work on that subject and provided the production database of the journal. The author thanks Denise Pumain for helping with expert geographical knowledge in the naming of communities. The author also thanks two anonymous reviewers which comments were of great value for the paper.

\end{document}